\documentclass[twocolumn,showpacs,preprintnumbers,amsmath,amssymb,subfig]{revtex4}

\usepackage{graphicx}
\usepackage{dcolumn}
\usepackage{bm}

\begin{document}

\preprint{}

\title{Photoluminescence and spectral switching of single CdSe/ZnS colloidal nanocrystals in poly(methyl methacrylate)}

\author{Y.M.~Shen}
\email{ymshen@ucsd.edu}
\author{L.~Pang}
\author{Y.~Fainman}%
 \homepage{http://emerald.ucsd.edu}
\affiliation{%
Electrical and Computer Science Engineering Department,\\  University of California, San Diego\\}%

\author{M.~Griswold}
\author{Sen~Yang}
\author{L.V.~Butov}
 \homepage{http://physics.ucsd.edu/~lvbutov}
\author{L.J.~Sham}%
\homepage{http://physics.ucsd.edu/~ljssst/ljs.html}
 \affiliation{Department of Physics, University of California, San Diego\\
}%

\date{\today}

\begin{abstract}
Emission from single CdSe nanocrystals in PMMA was investigated. A
fraction of the nanocrystals exhibiting switching between two energy
states, which have similar total intensities, but distinctly
different spectra were observed. We  found that the spectral shift
characteristic frequency increases with the pump power. By using the
dynamic shift in the spectral position of emission peaks, we were
able to correlate peaks from the same nanocrystal. The measured
correlation is consistent with assignment of low energy lines to
phonon replicas.

\end{abstract}

\maketitle

\section{\label{sec:level1}INTRODUCTION\protect\\    }

Semiconductor quantum dots (QDs) have attracted much interest due to
their properties of nanoscale quantum confinement that posses great
promise for numerous optoelectronics and photonics applications
including displays \cite{Colvin1994}, and biology
\cite{Bruchez1998}. While there are many possible fabrication
methods, semiconductor QDs formed in colloidal solution, called
nanocrystal (NC), via chemical synthesis have been shown to be a
promising route to QD realization
\cite{Wang1987,Murray1993,Hines1996}. In colloidal solutions,
however, NC's are suspended in a solvent, making them less practical
for fabrication and integration of photonic and optoelectronic
devices. The introduction of such NC's into a solid-state matrix,
therefore, is of great interest for numerous applications. We have
succeeded in incorporating the NC's into Polymethyl
methacrylate(PMMA) matrix \cite{lin1, lin2}. The sensitivity of this
composite to electron beams makes it attractive in the fabrication
of photonic devices. In this article, we particularly studied the
spectra of a single NC in PMMA. It's useful for the future
application of a single NC in PMMA photonics devices.

   Novel information, concealed by the inhomogeneously-broadened
photoluminescence (PL) lines, has been revealed by a single QD
spectroscopy. Two of the observations include fluctuations in PL
intensity and peak energy with time-- referred to as fluorescence
intermittency and spectrum diffusion, respectively. Both phenomena
have been observed in colloidal NC and self-assembled QD's.

Fluorescence intermittency has been well studied in colloidal NC's
 \cite{Nirmal,Empedocles1996,SA1997}, silicon
  nanocrystals \cite{Ilya2005},
and some of the self-assembled QD's, including CdSe/ZnSe
\cite{Turck}, and Ga$_{0.6}$In$_{0.4}$As quantum dots in GaAs
\cite{Panev2001}. The intermittency is reported to display a
telegraph signal \cite{Pistol1999,Pistol2001,Panev2001} where the
switching rate can be very slow, with typical time scales of seconds
or even minutes. A model based on photoionization has been used to
explain the switching for colloidal dots \cite{Empedocles1996},
while for expitaxially grown dots, a model involving a mobile
photoactivated defect has been proposed \cite{Pistol1999}.

Spectrum diffusion, especially random switching between discrete
levels, has been observed and well studied in some self-assembled
systems: InAs/GaAs QD's \cite{Landin}. It has also been found when
colloidal nanocrystals were deposited on gold substrates
\cite{Shimizu2002}. In this paper, we show that colloidal CdSe
nanocrystals in PMMA also exhibit similar behavior, and we
explicitly quantify the dependence on excitation power. Also a
jitter of the emission energies of the NC's was observed. The jitter
is characteristic for each NC and allow us to identify phonon
replicas for single NC's.

\section{\label{sec:level1}EXPERIMENT\protect\\  }

\input{epsf}

  The sample was prepared by using the colloidal CdSe/ZnS core shell NC's (Evident Technologies). The CdSe core is about 5~nm in diameter
capped by a few-monolayer ZnS shell to increase the quantum yield. A
PMMA layer of about 200~nm thick was spin coated on a quartz
substrate, followed by a spin coat of NC's in toluene, leading to
CdSe/ZnS NC's distributed in a PMMA matrix.  The final density was
chosen to be less than one dot per
 $\mu$m$^{2}$.

\begin{figure*}
\epsfxsize=7in \epsfysize=5in \epsffile{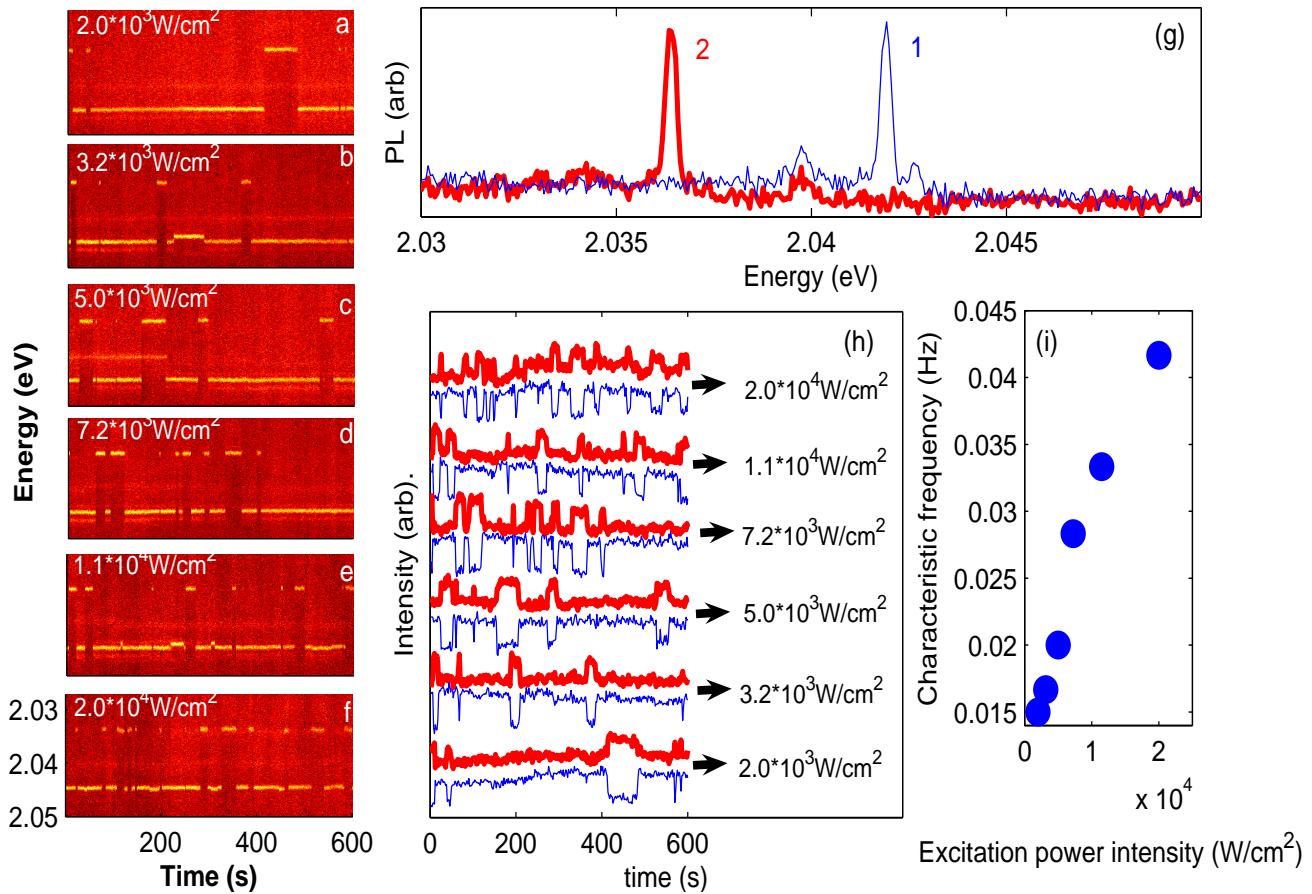}
\caption{\label{fig:wide}(Color online) Spectra from a typical
switching NC's, recorded at six different excitation powers
(2.0,3.2,5.0,7.2,11,20*10$^{3}$W/cm$^{2}$) are shown in order in
(a)-(f). The integration time 2~s is used for this measurement. The
brightness represents the intensity, the horizontal axis is the time
and the vertical axis is emission energy. (g) is the integration for
26 measurements (each measurement has 2~s integration time) when the
NC's stay in the same state. The energy difference between the two
states is 5.6~meV. (h) intensity vs. time plot for Fig.1(a)-1(f).
The (blue) thin line is for the high energy state. (i) The
characteristic switching frequency vs excitation power. }
\end{figure*}

For PL measurements, the sample was placed into a continuous flow
cryostat, and the temperature was kept constant at 7~K by
controlling the flow of liquid Helium.  A diode pumped solid state
laser operating at 532~nm was used for excitation, with power
densities typically between 1-20~kW/cm$^{2}$, and was varied using
neutral density filters. The luminescence from the sample was
collected through an optical microscope objective (50x, numerical
aperture 0.55) with a long working distance. This collected light
passed through a notch filter, dispersed in a monochromator, and was
detected using a liquid-nitrogen-cooled charge-coupled device (CCD)
camera. The spectral resolution of the setup was 200~$\mu$eV.
Spectra from a single NC were monitored continuously with an
integration time varied between 500~ms and 30~s depending on the
excitation power density and NC brightness.

\begin{figure}
\epsfxsize=3.5in \epsfysize=3in \epsffile{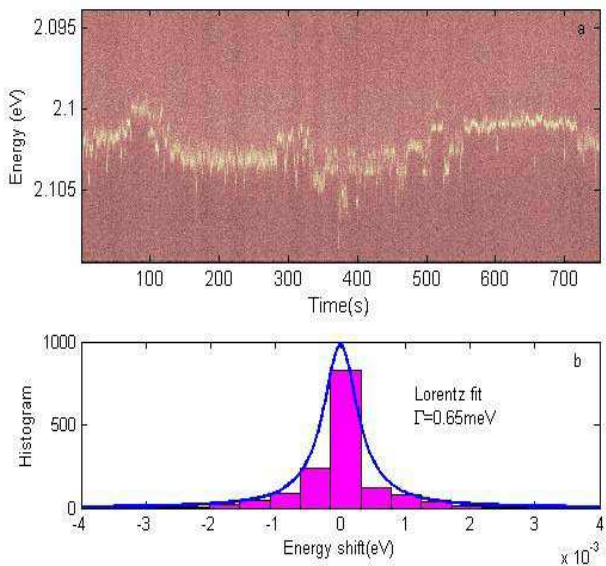}
\caption{\label{fig:epsart} (color online) (a) Spectra from another
NC were recorded continuously with an integration time of 0.5~s. The
emission peak randomly shifts with small freq. changes between
2.099~eV and 2.108~eV. (b)Energy shift from one measurement to the
next measurement is compared with a Lorentz distribution.}
\end{figure}

\section{\label{sec:level1}RESULTS\protect\\  }

The density of NC's was made so dilute that we typically observed
only a single NC in the interrogated spatial and spectral range. We
found and investigated several NC's switching between two states.
Spectra from a typical switching NC, recorded at six different
excitation powers, are shown in Fig.~1(a)-1(f).These spectra are
recorded continuously with an integration time of 2~s each. The
greyscale plot represents intensity, while the horizontal axis is
time, and the vertical axis is the emission energy. Fig.~1(g) shows
the typical spectrum for the two states. The red line(bold line) is
for the lower energy state, which we will call state 2, while the
blue line(thin line) is for the higher energy state which we call
state 1. The energy difference between the two states is 5.6~meV.
Fig.~1(h) is a plot of intensity vs. time for data taken from
Fig.~1(a)-1(f). Again, the red line(bold line) is state 2 and the
blue line(thin line) is state 1. Fig.~1(i) is the plot of the
characteristic frequency (inverse average time) vs excitation power.
From Fig.~1(g) and 1(h) we see that there is no sizable change in
the integrated PL intensity for the two states. The excitation power
density is of the order of 10~kW/cm$^{2}$, which is similar to that
used for measurements with InP QD's \cite{Panev2004}.

Figure 1 shows that the switching frequency increases with
increasing excitation power density. Since the widths of both state
plateaus decrease with increasing excitation power, we can conclude
that the transitions from the two states are light induced.

Some NC's were observed to operate outside of the two-state
description. We found NC's switching between three levels or more.

We also observed some NC's exhibiting random energy shifts, similar
to behavior seen for those NC's without PMMA \cite{Empedocles1996}.
In Fig.~2, we analyze a series of 1500 spectra taken in intervals of
0.5 s. The spectrum peak is seen to wander randomly. The energy
change for the two neighbor spectrum can be approximated by a
Lorentz[see Figs. 2(b)]

Some NC's can be permanently quenched after a period of time. These
NC's suddenly become dark after continuous illumination and do not
recover in the measurement time (the longest time we monitored was 2
hours ).

Figure~3 (a-c) shows that in some NC's, higher laser intensity not
only increases the switching frequency, but also makes the states
moving randomly. In addition, the figures show that there is an
additional peak in lower energy side with the energy difference
22.7~meV. The energy of the lines are shifting in step. Therefore,
we infer that these lines come from a single NC. The right side of
Figure~3(a-c) shows the normalized spectrum respectively. We
attribute the side peaks to phonon replicas since the ratio of its
intensity to the exciton intensity remains constant when the pump
intensity changes.

In order to observe more spectral lines,  we integrated over
30~second for each frame in figure 4. We record 200 measurements
continuously with constant temperature and excitation power. Again
the energy shifts of the lines are correlated, and therefore, we
conclude that these lines come from a single NC. The energy
difference between the main line and the satellite lines are
22.7~meV and 25.8~meV, respectively. Also we can see more lines on
the lower energy side. The energy differences with the main line are
45.4~meV,48.5~meV respectively. We identified them as the lines from
surface optical(SO) phonons, longitudinal optical (LO) phonons, two
SO phonons, and one SO plus one LO phonons, see section IV.

\begin{figure}
\epsfxsize=3.5in \epsfysize=3in
\epsffile{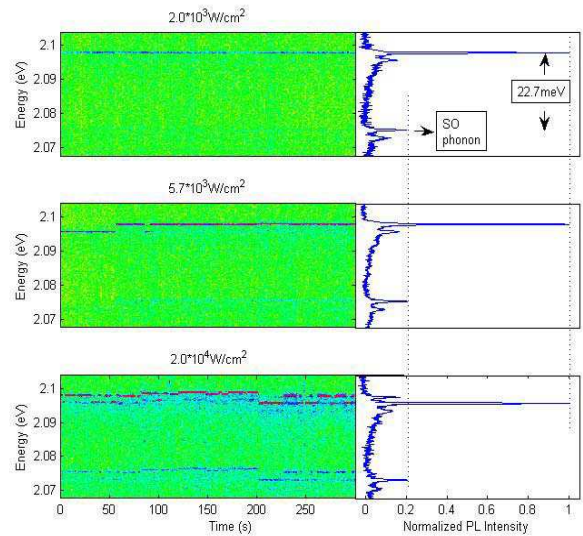}
\caption{\label{fig:epsart}(Color online)Spectra from a third NC
were recorded continuously with an integration time of 2 s each.
Figure 3(a-c) shows in some NC's, higher laser intensity not only
increases the switching frequency, but also makes the states moving
randomly. In addition, the figures show there is an additional peak
in lower energy side.  The energy difference with the exciton peak
is 22.7 meV.}
\end{figure}

\begin{figure}
\epsfxsize=3.5in \epsfysize=3in \epsffile{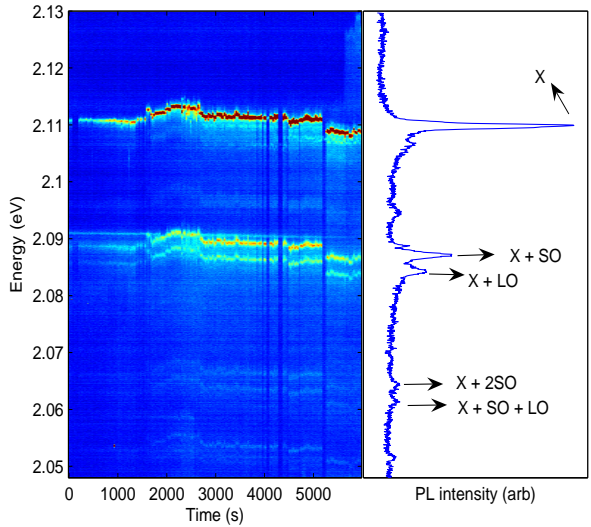}
\caption{\label{fig:epsart}(Color online) Spectra from a fourth NC
were recorded continuously with an integration time of 30 s each and
constant temperature and excitation power. We used a 30s integration
time for each frame to see the higher energy states which are
usually weaker than the exciton peak.  Spectrum jitter allows us to
group spectral transitions from individual dots. The energy
differences between the main line and the satellite lines are 22.7,
25.8, 45.4, 48.5 meV}
\end{figure}

\section{\label{sec:level1}DISCUSSION\protect\\  }
The blinking behavior of CdSe/ZnS NC's has attracted much scientific
attention \cite{Nirmal,Empedocles1996,SA1997}. Through this body of
work, it was found that the statistics of fluorescence intermittency
in single CdSe NC's exhibit a power-law distribution in the
histogram of on and off times, i.e., the time period before the NC
turns from emitting to nonemitting (bright to dark) and vice versa.
Every NC shows a similar power-law behavior for the off-time
distribution regardless of temperature, excitation intensity,
surface morphology or size \cite{Shimizu2001}. In our experiment,
the blinking behavior does not play a major role, which could be
caused by the surrounding PMMA matrix.  The spectral shift is
dominant most of the time.

Besides the normally observed random spectral shifts (Fig.~2a), we
have found large abrupt spectral jumps with more than one energy
change. The two emission peaks in Fig.~1(g) seldom emit together
within our integration time, which excludes the possibility of
biexciton emission and optical phonon progressions. If a biexcition
were created, we would observe both lines simultaneously within our
integration time as the biexcition relaxes to the excition
radiatively, followed by exciton emission. The same observation
applies to optical phonon replicas.

In 2002, Bawendi's group also found a spectral shift between two
positions when they put CdSe/ZnS NC on rough metal surface. The
emission energy fluctuates with a spacing of 15-25~meV
\cite{Shimizu2002}. They suggested that the observed emission shifts
were caused by neutral (X) and charged (X') exciton emission from a
single NC. However theoretical calculation indicated that the
charged NC could emit 25~meV to the red of the neutral exciton
emission \cite{Franceschetti2000}. Both are different from our
result since our energy spacing between two positions is 5.6~meV or
smaller.

Similar spectral switching behavior has also been observed in InP
quantum dots in Ga$_{x}$In$_{1-x}$P matrix grown by metal-organic
vapor-phase epitaxy. They found the quantum dot spectrum switches
between two states which have the same integrated PL, similar to our
result. The difference is that each state has multiple lines in
their experiment \cite{Panev2004}.

We conclude that the spectral shift is a universal behavior for all
quantum dots of III-V and II-IV compound self assembled QD's and
colloidal nanocrystals.  A new theory is needed to explain this
behavior.

The observed peaks in lower energy can originate from surface
optical (SO) phonons or longitudinal optical (LO) phonons in the
NC's. The energies of the SO modes, $\Omega_{SO}$, are determined by
the energy of the TO phonons, $\Omega_{TO}$, in CdSe NC's, the shape
of the NC's , and the dielectric constants of the core and
surrounding medium. For spherical NC's the classical dispersion
relation for the interface phonons can be presented by the following
equation \cite{PhysRevB.56.7491}:
\begin{equation}
\Omega_{SO}=\Omega_{TO}[\frac{\varepsilon_{0}l+\varepsilon_{M}(l+1)}{\varepsilon_{\infty}l+\varepsilon_{M}(l+1)}]^{1/2}
\end{equation}
where $\varepsilon_{0}$ and $\varepsilon_{\infty}$ are the static
and high-frequency dielectric constants of the bulk CdSe and
$\varepsilon_{M}$ is the static dielectric constant of the
surrounding medium. Solely SO-phonons with $\ell$=even integers are
allowed (SO-phonons with $\ell$=odd integers are forbidden). By
using the bulk CdSe values of $\Omega_{TO}=167.5~cm^{-1}$,
$\varepsilon_{0}=9.3$, $\varepsilon_{\infty}=6.1$ \cite{landolt} and
the dielectric constant of hexagonal ZnS $\varepsilon_{M}=8.3$, we
have calculated for the lowest ($\ell$=2) and the highest
($\ell$$\rightarrow\infty$) modes of spherical CdSe NC's values of
$\Omega_{SO}^{l=2}=180~cm^{-1} (22.4~meV)$ and
$\Omega_{SO}^{l=\infty}=185~cm^{-1} (23.0~meV)$, respectively. The
calculated energy is close to the experiment value (22.7~meV).

LO phone energy 25.8~meV is smaller than the bulk value of 26.1~meV
\cite{CdSeBulk}, and is in good agreement with the values expected
from theory for a 25A radius CdSe nano-crystal, which are in the
range 23.8-26.0~meV for the first four LO phonons.
\cite{PhysRevB.57.4664}

The Huang-Rhys factors (electron-phonon coupling strength) can be
roughly estimated from ratio between integrated intensities of the
2SO and SO lines. We have obtain value of $S\approx0.2$.

\section{\label{sec:level1}CONCLUSIONS\protect\\  }
We have investigated random switching between two states in the PL
from CdSe/ZnS colloidal nanocrystals in PMMA. The commonly observed
blinking behavior was suppressed.  The spectrum switching behaviors
are dominant and some of them are similar to what has been observed
in III-V quantum dots. To explain these observations new models
going beyond those established in the literature need to be
developed. We also found the random energy shift which follows
approximately a Lorentzian distribution. Identical jitter patterns
allow the unambiguous identification of the emission spectra of
single NC's. By means of excitation density dependent spectroscopy
the lines are identified as phonon replica of the exciton transition
involving local SO-phonon and LO-phonon modes of the NC. Huang-Rhys
factors of $S\approx0.2$ are observed.

\begin{acknowledgments}
We thank Kevin Tetz and Robert Saperstein for fruitful discussion.
This work was funded by DARPA, NSF and AFOSR.
\end{acknowledgments}

\newpage 
\bibliography{apssamp}

\begin{thebibliography}{24}
\expandafter\ifx\csname natexlab\endcsname\relax\def\natexlab#1{#1}\fi
\expandafter\ifx\csname bibnamefont\endcsname\relax
  \def\bibnamefont#1{#1}\fi
\expandafter\ifx\csname bibfnamefont\endcsname\relax
  \def\bibfnamefont#1{#1}\fi
\expandafter\ifx\csname citenamefont\endcsname\relax
  \def\citenamefont#1{#1}\fi
\expandafter\ifx\csname url\endcsname\relax
  \def\url#1{\texttt{#1}}\fi
\expandafter\ifx\csname urlprefix\endcsname\relax\def\urlprefix{URL }\fi
\providecommand{\bibinfo}[2]{#2}
\providecommand{\eprint}[2][]{\url{#2}}

\bibitem[{\citenamefont{Colvin et~al.}(1994)\citenamefont{Colvin, Schlamp, and
  Alivisatos}}]{Colvin1994}
\bibinfo{author}{\bibfnamefont{V.}~\bibnamefont{Colvin}},
  \bibinfo{author}{\bibfnamefont{M.}~\bibnamefont{Schlamp}}, \bibnamefont{and}
  \bibinfo{author}{\bibfnamefont{A.}~\bibnamefont{Alivisatos}},
  \bibinfo{journal}{Nautre(London)} \textbf{\bibinfo{volume}{370}},
  \bibinfo{pages}{354} (\bibinfo{year}{1994}).

\bibitem[{\citenamefont{Bruchez et~al.}(1998)\citenamefont{Bruchez, Moronne,
  Gin, Weiss, and Alivisatos}}]{Bruchez1998}
\bibinfo{author}{\bibfnamefont{M.}~\bibnamefont{Bruchez}},
  \bibinfo{author}{\bibfnamefont{M.}~\bibnamefont{Moronne}},
  \bibinfo{author}{\bibfnamefont{P.}~\bibnamefont{Gin}},
  \bibinfo{author}{\bibfnamefont{S.}~\bibnamefont{Weiss}}, \bibnamefont{and}
  \bibinfo{author}{\bibfnamefont{A.~P.} \bibnamefont{Alivisatos}},
  \bibinfo{journal}{Science} \textbf{\bibinfo{volume}{281}},
  \bibinfo{pages}{2013} (\bibinfo{year}{1998}).

\bibitem[{\citenamefont{Wang et~al.}(1987)\citenamefont{Wang, Suna, Mahler, and
  Kasowski}}]{Wang1987}
\bibinfo{author}{\bibfnamefont{Y.}~\bibnamefont{Wang}},
  \bibinfo{author}{\bibfnamefont{A.}~\bibnamefont{Suna}},
  \bibinfo{author}{\bibfnamefont{W.}~\bibnamefont{Mahler}}, \bibnamefont{and}
  \bibinfo{author}{\bibfnamefont{R.}~\bibnamefont{Kasowski}},
  \bibinfo{journal}{J. Chem. Phys.} \textbf{\bibinfo{volume}{87}},
  \bibinfo{pages}{7315} (\bibinfo{year}{1987}).

\bibitem[{\citenamefont{Murray et~al.}(1993)\citenamefont{Murray, Norris, and
  Bawendi}}]{Murray1993}
\bibinfo{author}{\bibfnamefont{C.~B.} \bibnamefont{Murray}},
  \bibinfo{author}{\bibfnamefont{D.~J.} \bibnamefont{Norris}},
  \bibnamefont{and} \bibinfo{author}{\bibfnamefont{M.~G.}
  \bibnamefont{Bawendi}}, \bibinfo{journal}{J. Am. Chem. Soc}
  \textbf{\bibinfo{volume}{115}}, \bibinfo{pages}{8706} (\bibinfo{year}{1993}).

\bibitem[{\citenamefont{Hines and Guyot-Sionnest}(1996)}]{Hines1996}
\bibinfo{author}{\bibfnamefont{M.~A.} \bibnamefont{Hines}} \bibnamefont{and}
  \bibinfo{author}{\bibfnamefont{P.}~\bibnamefont{Guyot-Sionnest}},
  \bibinfo{journal}{J. Phys. Chem.} \textbf{\bibinfo{volume}{100}},
  \bibinfo{pages}{468} (\bibinfo{year}{1996}).

\bibitem[{\citenamefont{Pang et~al.}(2005{\natexlab{a}})\citenamefont{Pang,
  Shen, Tetz, and Fainman}}]{lin1}
\bibinfo{author}{\bibfnamefont{L.}~\bibnamefont{Pang}},
  \bibinfo{author}{\bibfnamefont{Y.}~\bibnamefont{Shen}},
  \bibinfo{author}{\bibfnamefont{K.}~\bibnamefont{Tetz}}, \bibnamefont{and}
  \bibinfo{author}{\bibfnamefont{Y.}~\bibnamefont{Fainman}},
  \bibinfo{journal}{Opt. Express} \textbf{\bibinfo{volume}{13}},
  \bibinfo{pages}{44} (\bibinfo{year}{2005}{\natexlab{a}}).

\bibitem[{\citenamefont{Pang et~al.}(2005{\natexlab{b}})\citenamefont{Pang,
  Tetz, Shen, Chen, and Fainman}}]{lin2}
\bibinfo{author}{\bibfnamefont{L.}~\bibnamefont{Pang}},
  \bibinfo{author}{\bibfnamefont{K.}~\bibnamefont{Tetz}},
  \bibinfo{author}{\bibfnamefont{Y.}~\bibnamefont{Shen}},
  \bibinfo{author}{\bibfnamefont{C.}~\bibnamefont{Chen}}, \bibnamefont{and}
  \bibinfo{author}{\bibfnamefont{Y.}~\bibnamefont{Fainman}},
  \bibinfo{journal}{J. Vac. Sci. Technol. B} \textbf{\bibinfo{volume}{23}},
  \bibinfo{pages}{2413} (\bibinfo{year}{2005}{\natexlab{b}}).

\bibitem[{\citenamefont{Nirmal et~al.}(1996)\citenamefont{Nirmal, Dabbousi,
  Bawendi, Macklin, Trautman, Harris, and Brus}}]{Nirmal}
\bibinfo{author}{\bibfnamefont{M.}~\bibnamefont{Nirmal}},
  \bibinfo{author}{\bibfnamefont{B.~O.} \bibnamefont{Dabbousi}},
  \bibinfo{author}{\bibfnamefont{M.~G.} \bibnamefont{Bawendi}},
  \bibinfo{author}{\bibfnamefont{J.~J.} \bibnamefont{Macklin}},
  \bibinfo{author}{\bibfnamefont{J.~K.} \bibnamefont{Trautman}},
  \bibinfo{author}{\bibfnamefont{T.~D.} \bibnamefont{Harris}},
  \bibnamefont{and} \bibinfo{author}{\bibfnamefont{L.~E.} \bibnamefont{Brus}},
  \bibinfo{journal}{Nature} \textbf{\bibinfo{volume}{383}},
  \bibinfo{pages}{802} (\bibinfo{year}{1996}).

\bibitem[{\citenamefont{Empedocles et~al.}(1996)\citenamefont{Empedocles,
  Norris, and Bawendi}}]{Empedocles1996}
\bibinfo{author}{\bibfnamefont{S.~A.} \bibnamefont{Empedocles}},
  \bibinfo{author}{\bibfnamefont{D.~J.} \bibnamefont{Norris}},
  \bibnamefont{and} \bibinfo{author}{\bibfnamefont{M.~G.}
  \bibnamefont{Bawendi}}, \bibinfo{journal}{Phys.\ Rev.B}
  \textbf{\bibinfo{volume}{77}}, \bibinfo{pages}{3873} (\bibinfo{year}{1996}).

\bibitem[{\citenamefont{Empedocles and Bawendi}(1997)}]{SA1997}
\bibinfo{author}{\bibfnamefont{S.~A.} \bibnamefont{Empedocles}}
  \bibnamefont{and} \bibinfo{author}{\bibfnamefont{M.~G.}
  \bibnamefont{Bawendi}}, \bibinfo{journal}{Science}
  \textbf{\bibinfo{volume}{278}}, \bibinfo{pages}{2114} (\bibinfo{year}{1997}).

\bibitem[{\citenamefont{Sychugov et~al.}(2005)\citenamefont{Sychugov, Juhasz,
  and Linnros}}]{Ilya2005}
\bibinfo{author}{\bibfnamefont{I.}~\bibnamefont{Sychugov}},
  \bibinfo{author}{\bibfnamefont{R.}~\bibnamefont{Juhasz}}, \bibnamefont{and}
  \bibinfo{author}{\bibfnamefont{J.}~\bibnamefont{Linnros}},
  \bibinfo{journal}{Phys. Rev. B} \textbf{\bibinfo{volume}{71}},
  \bibinfo{pages}{115331} (\bibinfo{year}{2005}).

\bibitem[{\citenamefont{Turck et~al.}(2000)\citenamefont{Turck, Rodt, Stier,
  Heitz, Engelhardt, Pohl, Bimberg, and Steingruber}}]{Turck}
\bibinfo{author}{\bibfnamefont{V.}~\bibnamefont{Turck}},
  \bibinfo{author}{\bibfnamefont{S.}~\bibnamefont{Rodt}},
  \bibinfo{author}{\bibfnamefont{O.}~\bibnamefont{Stier}},
  \bibinfo{author}{\bibfnamefont{R.}~\bibnamefont{Heitz}},
  \bibinfo{author}{\bibfnamefont{R.}~\bibnamefont{Engelhardt}},
  \bibinfo{author}{\bibfnamefont{U.~W.} \bibnamefont{Pohl}},
  \bibinfo{author}{\bibfnamefont{D.}~\bibnamefont{Bimberg}}, \bibnamefont{and}
  \bibinfo{author}{\bibfnamefont{R.}~\bibnamefont{Steingruber}},
  \bibinfo{journal}{Phys.\ Rev.B} \textbf{\bibinfo{volume}{61}},
  \bibinfo{pages}{9944} (\bibinfo{year}{2000}).

\bibitem[{\citenamefont{Panev et~al.}(2001)\citenamefont{Panev, Pistol,
  Zwiller, Samuelson, Jiang, Xu, and Wang}}]{Panev2001}
\bibinfo{author}{\bibfnamefont{N.}~\bibnamefont{Panev}},
  \bibinfo{author}{\bibfnamefont{M.~E.} \bibnamefont{Pistol}},
  \bibinfo{author}{\bibfnamefont{V.}~\bibnamefont{Zwiller}},
  \bibinfo{author}{\bibfnamefont{L.}~\bibnamefont{Samuelson}},
  \bibinfo{author}{\bibfnamefont{W.}~\bibnamefont{Jiang}},
  \bibinfo{author}{\bibfnamefont{B.}~\bibnamefont{Xu}}, \bibnamefont{and}
  \bibinfo{author}{\bibfnamefont{Z.}~\bibnamefont{Wang}},
  \bibinfo{journal}{Phys.\ Rev.B} \textbf{\bibinfo{volume}{64}},
  \bibinfo{pages}{045317} (\bibinfo{year}{2001}).

\bibitem[{\citenamefont{Pistol et~al.}(1999)\citenamefont{Pistol, Castrillo,
  Hessman, Prieto, and Samuelson}}]{Pistol1999}
\bibinfo{author}{\bibfnamefont{M.~E.} \bibnamefont{Pistol}},
  \bibinfo{author}{\bibfnamefont{P.}~\bibnamefont{Castrillo}},
  \bibinfo{author}{\bibfnamefont{D.}~\bibnamefont{Hessman}},
  \bibinfo{author}{\bibfnamefont{J.~A.} \bibnamefont{Prieto}},
  \bibnamefont{and}
  \bibinfo{author}{\bibfnamefont{L.}~\bibnamefont{Samuelson}},
  \bibinfo{journal}{Phys.\ Rev.B} \textbf{\bibinfo{volume}{59}},
  \bibinfo{pages}{10725} (\bibinfo{year}{1999}).

\bibitem[{\citenamefont{Pistol}(2001)}]{Pistol2001}
\bibinfo{author}{\bibfnamefont{M.~E.} \bibnamefont{Pistol}},
  \bibinfo{journal}{Phys.\ Rev.B} \textbf{\bibinfo{volume}{63}},
  \bibinfo{pages}{113306} (\bibinfo{year}{2001}).

\bibitem[{\citenamefont{Landin et~al.}(1998)\citenamefont{Landin, Miller,
  Pistol, Pryor, and Samuelson}}]{Landin}
\bibinfo{author}{\bibfnamefont{L.}~\bibnamefont{Landin}},
  \bibinfo{author}{\bibfnamefont{M.~S.} \bibnamefont{Miller}},
  \bibinfo{author}{\bibfnamefont{M.~E.} \bibnamefont{Pistol}},
  \bibinfo{author}{\bibfnamefont{C.~E.} \bibnamefont{Pryor}}, \bibnamefont{and}
  \bibinfo{author}{\bibfnamefont{L.}~\bibnamefont{Samuelson}},
  \bibinfo{journal}{Science} \textbf{\bibinfo{volume}{80}},
  \bibinfo{pages}{262} (\bibinfo{year}{1998}).

\bibitem[{\citenamefont{Shimizu et~al.}(2002)\citenamefont{Shimizu, Woo,
  Fisher, Eisler, and Bawendi}}]{Shimizu2002}
\bibinfo{author}{\bibfnamefont{K.~T.} \bibnamefont{Shimizu}},
  \bibinfo{author}{\bibfnamefont{W.~K.} \bibnamefont{Woo}},
  \bibinfo{author}{\bibfnamefont{B.~R.} \bibnamefont{Fisher}},
  \bibinfo{author}{\bibfnamefont{H.~J.} \bibnamefont{Eisler}},
  \bibnamefont{and} \bibinfo{author}{\bibfnamefont{M.~G.}
  \bibnamefont{Bawendi}}, \bibinfo{journal}{Phys.\ Rev.Lett}
  \textbf{\bibinfo{volume}{89}}, \bibinfo{pages}{117401}
  (\bibinfo{year}{2002}).

\bibitem[{\citenamefont{Panev et~al.}(2004)\citenamefont{Panev, Pistol,
  Persson, Seifert, and Samuelson}}]{Panev2004}
\bibinfo{author}{\bibfnamefont{N.}~\bibnamefont{Panev}},
  \bibinfo{author}{\bibfnamefont{M.~E.} \bibnamefont{Pistol}},
  \bibinfo{author}{\bibfnamefont{J.}~\bibnamefont{Persson}},
  \bibinfo{author}{\bibfnamefont{W.}~\bibnamefont{Seifert}}, \bibnamefont{and}
  \bibinfo{author}{\bibfnamefont{L.}~\bibnamefont{Samuelson}},
  \bibinfo{journal}{Phys.\ Rev.B} \textbf{\bibinfo{volume}{70}},
  \bibinfo{pages}{073309} (\bibinfo{year}{2004}).

\bibitem[{\citenamefont{Shimizu et~al.}(2001)\citenamefont{Shimizu, Neuhauser,
  Leatherdale, Empedocles, Woo, and Bawendi}}]{Shimizu2001}
\bibinfo{author}{\bibfnamefont{K.~T.} \bibnamefont{Shimizu}},
  \bibinfo{author}{\bibfnamefont{R.~G.} \bibnamefont{Neuhauser}},
  \bibinfo{author}{\bibfnamefont{C.~A.} \bibnamefont{Leatherdale}},
  \bibinfo{author}{\bibfnamefont{S.~A.} \bibnamefont{Empedocles}},
  \bibinfo{author}{\bibfnamefont{W.~K.} \bibnamefont{Woo}}, \bibnamefont{and}
  \bibinfo{author}{\bibfnamefont{M.~G.} \bibnamefont{Bawendi}},
  \bibinfo{journal}{Phys.\ Rev.B} \textbf{\bibinfo{volume}{63}},
  \bibinfo{pages}{205316} (\bibinfo{year}{2001}).

\bibitem[{\citenamefont{Franceschetti and Zunger}(2000)}]{Franceschetti2000}
\bibinfo{author}{\bibfnamefont{A.}~\bibnamefont{Franceschetti}}
  \bibnamefont{and} \bibinfo{author}{\bibfnamefont{A.}~\bibnamefont{Zunger}},
  \bibinfo{journal}{Phys.\ Rev.B} \textbf{\bibinfo{volume}{62}},
  \bibinfo{pages}{16287} (\bibinfo{year}{2000}).

\bibitem[{\citenamefont{Fedorov et~al.}(1997)\citenamefont{Fedorov, Baranov,
  and Inoue}}]{PhysRevB.56.7491}
\bibinfo{author}{\bibfnamefont{A.~V.} \bibnamefont{Fedorov}},
  \bibinfo{author}{\bibfnamefont{A.~V.} \bibnamefont{Baranov}},
  \bibnamefont{and} \bibinfo{author}{\bibfnamefont{K.}~\bibnamefont{Inoue}},
  \bibinfo{journal}{Phys. Rev. B} \textbf{\bibinfo{volume}{56}},
  \bibinfo{pages}{7491} (\bibinfo{year}{1997}).

\bibitem[{lan()}]{landolt}
\eprint{Semiconductors, edited by O. Madelung, W. von der Osten, and U.Rossler
  (Springer, Berlin, 1982), Landolt-Bernstein, New Series, Group III, Vol.
  22a.}

\bibitem[{CdS()}]{CdSeBulk}
\eprint{Numerical Data and Functional Relationship in Science and Technology,
  edited by K.H.Hellwege, Landolt-Bornstein, New Series, Group III, Vol. 17,
  Pt. B(Springer-Verlag, Berlin, 1982)}.

\bibitem[{\citenamefont{Trallero-Giner
  et~al.}(1998)\citenamefont{Trallero-Giner, Debernardi, Cardona,
  Men\'endez-Proup\'in, and Ekimov}}]{PhysRevB.57.4664}
\bibinfo{author}{\bibfnamefont{C.}~\bibnamefont{Trallero-Giner}},
  \bibinfo{author}{\bibfnamefont{A.}~\bibnamefont{Debernardi}},
  \bibinfo{author}{\bibfnamefont{M.}~\bibnamefont{Cardona}},
  \bibinfo{author}{\bibfnamefont{E.}~\bibnamefont{Men\'endez-Proup\'in}},
  \bibnamefont{and} \bibinfo{author}{\bibfnamefont{A.~I.}
  \bibnamefont{Ekimov}}, \bibinfo{journal}{Phys. Rev. B}
  \textbf{\bibinfo{volume}{57}}, \bibinfo{pages}{4664} (\bibinfo{year}{1998}).

\end{thebibliography}
\end{document}